\documentclass[10pt,a4paper]{article}
\usepackage[utf8]{inputenc}
\usepackage{amsmath}
\usepackage{amsfonts}
\usepackage{amssymb}
\usepackage{graphicx}
\usepackage[square,sort&compress,comma,numbers]{natbib}
\usepackage[separate-uncertainty = true,multi-part-units=single]{siunitx}
\usepackage{braket}
\usepackage{xcolor}
\usepackage{epstopdf}
\usepackage{color}

\usepackage[font=small,labelfont=bf]{caption}
\usepackage{authblk}

\title{Long-lived non-classical correlations for scalable quantum repeaters at room temperature}

\author[1]{Michael Zugenmaier}
\author[1]{Karsten B. Dideriksen}
\author[1]{Anders S. S\o rensen}
\author[1]{Boris Albrecht}
\author[1]{Eugene S. Polzik}
\affil[1]{Niels Bohr Institute, University of Copenhagen, DK-2100 Copenhagen, Denmark}
\date{}                     
\begin{document}
\maketitle

Heralded single-photon sources with on-demand readout are promising candidates for quantum repeaters enabling long-distance quantum communication. The need for scalability of such systems requires simple experimental solutions, thus favouring room-temperature systems.
For quantum repeater applications, long delays between heralding and single-photon readout are 
crucial. Until now, this has been prevented in room-temperature atomic systems by fast decoherence due to thermal motion.
Here we demonstrate efficient heralding and readout of single collective excitations created in warm caesium vapour. Using the principle of motional averaging we achieve a collective excitation lifetime of \SI{0.27(4)}{\milli\second}, two orders of magnitude larger than previously achieved for single excitations in room-temperature sources.
We experimentally verify non-classicality of the light-matter correlations by observing a violation of the Cauchy-Schwarz inequality with $R=\num{1.4(1)}>1$. 
Through spectral and temporal analysis we identify intrinsic four-wave mixing noise as the main contribution compromising single-photon operation of the source.

\pagebreak
Long-distance quantum communication beyond the limit of direct optical fiber transmission ($\sim$\SI{400}{\kilo\meter})\citep{Yin2016} requires a network of quantum repeater (QR) nodes. A QR increases the distance over which entanglement can be efficiently distributed by means of entanglement swapping \citep{Briegel1998}. Many attempts to realize such nodes are based on the so-called DLCZ protocol for atomic ensembles \citep{Duan2001}.

Key parameters for entanglement distribution in a QR network are the success rate of entanglement generation and the storage time in the elementary links of the network. For atomic ensemble-based QR nodes, this implies fast, high fidelity generation of heralded collective excitations and efficient retrieval after a controllable delay. The available storage time must exceed by far the average generation time as well as the end-to-end classical communication time of the network, although multiplexing of nodes can reduce the memory time requirement \citep{Sangouard2011}.

Since the first experimental realizations of the DLCZ protocol \citep{Kuzmich2003,VanderWal2003} more than a decade ago, frequent improvements in cold atomic ensembles have been reported \citep{Bao2012,Bimbard2014,Chen2006,Choi2010,Farrera2016,Inoue2009,Jiang2016,Laurat2006,Radnaev2010,Simon2007a,Yang2016,Zhao2008a} with memory times reaching \SI{0.22}{\second} \citep{Yang2016} and retrieval efficiencies up to \SI{84}{\percent} \citep{Simon2007a}. 
Progress has recently also been shown in solid-state systems, particularly in rare-earth-doped crystals \citep{Sinclair2014,Zhong2015,Kutluer2017,Laplane2017}. However, cryogenic cooling is required there. Room-temperature systems offer reliability and scalability, as they do not need cooling apparatus. Spin coherence with timescales of seconds in NV centers \citep{Maurer2012}, and minutes with atomic vapour in anti-relaxation-coated glass containers \citep{Balabas2010}, has been demonstrated at room temperature.
Still, coherent optical interaction with NV centers at room temperature remains a challenge \citep{Ghobadi2017}.
Broadband, short-lived quantum memories have been demonstrated in warm vapours \citep{Hosseini2011,Saunders2016}, but thermal atomic motion impedes long life spans of generated collective excitations or stored light \citep{Hockel2010,Ma2015,Phillips2008}. The utilization of buffer gas has allowed to extend the light storage duration at the few-photon level to \SI{20}{\micro\second} \citep{Namazi2017}. At the single-photon level, non-classical DLCZ-type correlations have been reported with buffer gas \citep{Eisaman2005,Manz2007,Bashkansky2012,Dou2017}, but with a lifetime limited to a few \SI{}{\micro\second}.
Anti-relaxation coating of the container walls has enabled continuous variable quantum memory of a few milliseconds \citep{Hammerer2010} and classical light storage up to \SI{150}{\milli\second} \citep{Katz2017}, but non-classical correlations for single excitations on such time scale remain to be observed.

In our work, we use the novel principle of motional averaging \citep{Borregaard2015} to demonstrate non-classical correlations with a lifetime extended by two orders of magnitude over previous realizations \citep{Dou2017,Bashkansky2012}. We confirm the non-classicality by observing the violation of the Cauchy-Schwarz inequality for field intensities \citep{Clauser1974}. The long memory time is verified by observation of a slowly decaying retrieval efficiency as the read-out delay is increased.

The readout fidelity in our system is limited by the excess noise in the read-out process leading to a high probability for detection events which do not originate from conversion of the collective excitation. We show that the noise contribution can be attributed mostly to four-wave mixing (FWM) \cite{Phillips2008,Lauk2013,Vurgaftman2013,Michelberger2015}, and we give examples of alternative excitation schemes for which the FWM noise can be avoided.
The motional averaging approach could serve as a solution toward the implemention of scalable quantum memories for applications such as spatially-multiplexed quantum networks, or deterministic single-photon sources for quantum information processing \citep{Aspuru-Guzik2012,Aaronson2011,Tillmann2013}.
To the best of our knowledge, it constitutes the only viable solution to a room-temperature QR without any need for cooling.

\pagebreak
\paragraph{Experimental setup.}

\begin{figure}[!ht]
	\centering
	\includegraphics[scale=0.55]{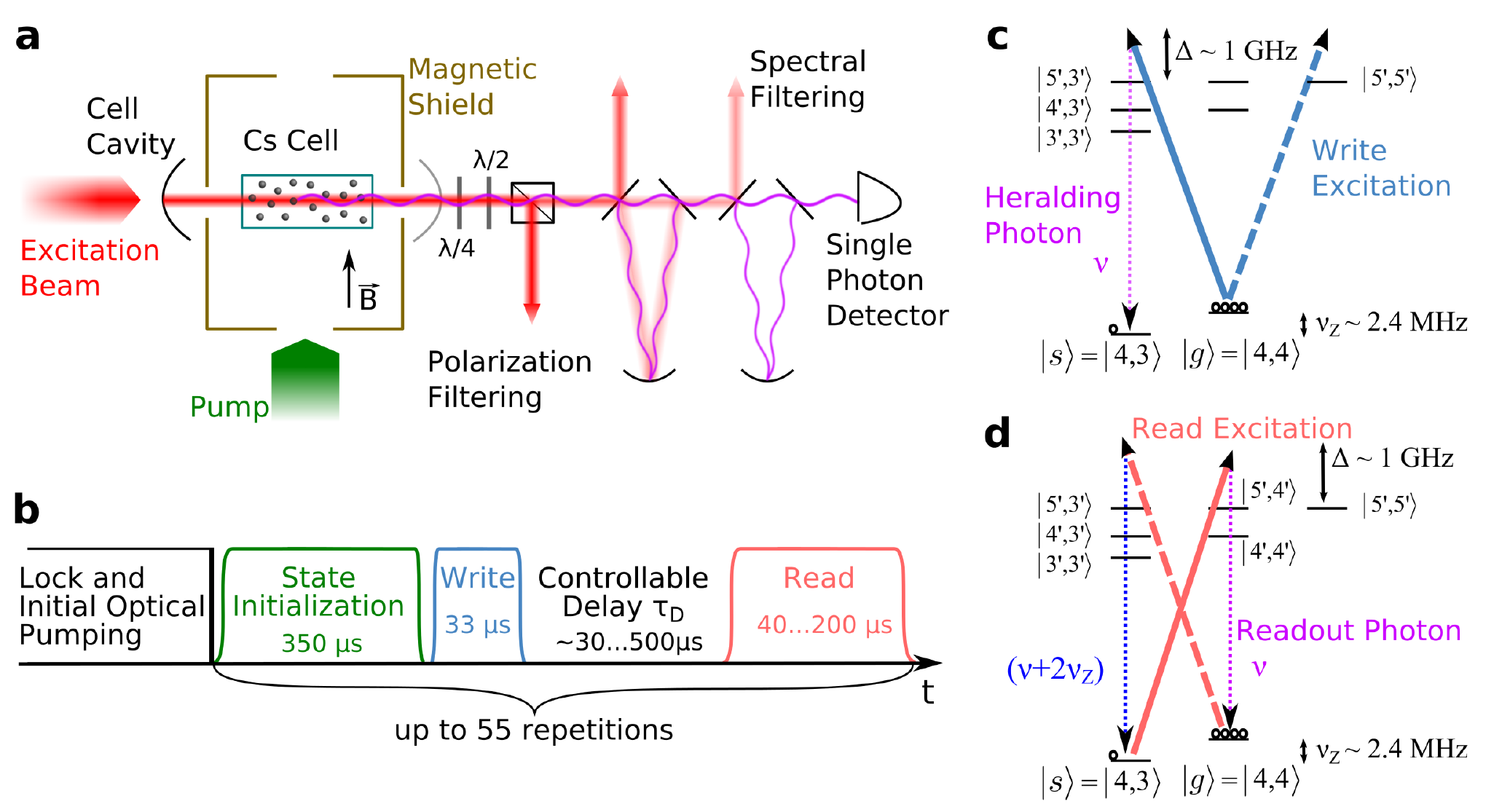} 
	\caption{\textbf{Experimental set-up and level scheme} (only main elements are shown).
		\textbf{a}, The caesium vapour cell is aligned with the Gaussian mode of the cell cavity. Optical pumping along the homogeneous magnetic field $\vec{\mathrm{B}}$ prepares the atoms in the $\ket{g}$ state. The excitation beam is coupled into the cell cavity. Forward-scattered photons passing through the two filtering cavities are detected with a single-photon detector, while the excitation light is filtered out by the cavities and a Glan-Thompson polarizer.
		\textbf{b}, The experimental sequence includes up to 55 cycles comprising pump, write and read pulses after each cavity locking period. The delay is defined as the difference between the on (off) times at half maximum for the read (write) pulses. 
		\textbf{c}, The write excitation pulse (solid line) acts on the atomic ensemble initially prepared in $\ket{g}$ and creates a single collective excitation in the Zeeman level $\ket{s}$. It also couples to the $\ket{F^\prime = 5, m^\prime_F = 5}$ excited state (dashed line) which has a slight influence on the collective excitation state.
		\textbf{d}, The read excitation pulse converts the collective excitation back into a photon. Simultaneously it acts on the initial state $\ket{g}$ (dashed line) creating more excitations, thus leading to spurious FWM noise.}
	\label{fig:ExpSetup}
\end{figure}

A vapour cell filled with caesium atoms, placed in a homogeneous magnetic field, is the basis for our experiment (fig.\ \ref{fig:ExpSetup}a). Paraffin coating of the cell walls preserves the atomic spin coherence upon hundreds of wall collisions.
The cell is aligned within a low finesse ($\mathcal{F}\approx 18$) asymmetric cell cavity, enhancing light-atom interaction. The light leaving the cell cavity passes through polarization and spectral filtering stages before detection by a single-photon counter (see Methods).
We initialize the caesium atoms via optical pumping into $\ket{g}\equiv\ket{F=4,m_F=4}$. A far-detuned, weak excitation pulse, linearly polarized perpendicular to the magnetic field, randomly scatters a photon via spontaneous Raman scattering (fig.\ \ref{fig:ExpSetup}c).
We herald the creation of a long-lived symmetric Dicke state \citep{Dicke1954} in $\ket{s}\equiv\ket{F=4,m_F=3}$ upon detection of such a photon scattered into the cell cavity mode. Since the transverse Gaussian profile of the cavity mode is narrower than the cell width, such detection events tend to be associated with asymmetric collective excitations distributed only on the atoms inside the beam at the time of detection, and consequently having a very limited lifetime. We overcome this by using motional averaging \citep{Borregaard2015}, extending the duration of the single photon wave packet, thus allowing for the atoms to cross the excitation beam several times. 
This is achieved by a narrow-band spectral filter, consisting of two optical cavities.  The spectral filter adds a random delay to the heralding photon, thus erasing "which path" information and ensuring that a detection event is equally likely to originate from any of the atoms, resulting in a long-lived symmetric collective excitation. The other purpose of the filtering cavities is to separate the excitation light from the scattered photon.

After a controllable delay $\tau_D$, the collective excitation is converted into a readout photon by a second, far-detuned pulse (fig.\ \ref{fig:ExpSetup}d).
Creating the collective excitation between Zeeman levels allows us to profit from their long coherence times.
However, the small Zeeman splitting of $\nu_Z = 2.4\,$\SI{}{\mega\hertz} presents a challenge for filtering out the excitation light. With our setup we achieve a suppression of the excitation light relative to the desired photon transmission by nine orders of magnitude. The read excitation light is chosen such that the readout photon is similar to the heralding photon in frequency and polarization. Thus only a single filtering and detection setup is required for both heralding and readout.

We start our experimental sequence by locking all cavities and initially optically pumping the atoms (fig.\ \ref{fig:ExpSetup}b). The following cycle comprising optical pumping for state re-initialization followed by write and read excitation pulses with controllable delay, is repeated up to \num{55} times before the sequence restarts, resulting in an average experimental repetition rate of up to \SI{1}{\kilo\hertz}.

\begin{figure}[ht!]
	\centering
	\includegraphics[width=0.8\textwidth]{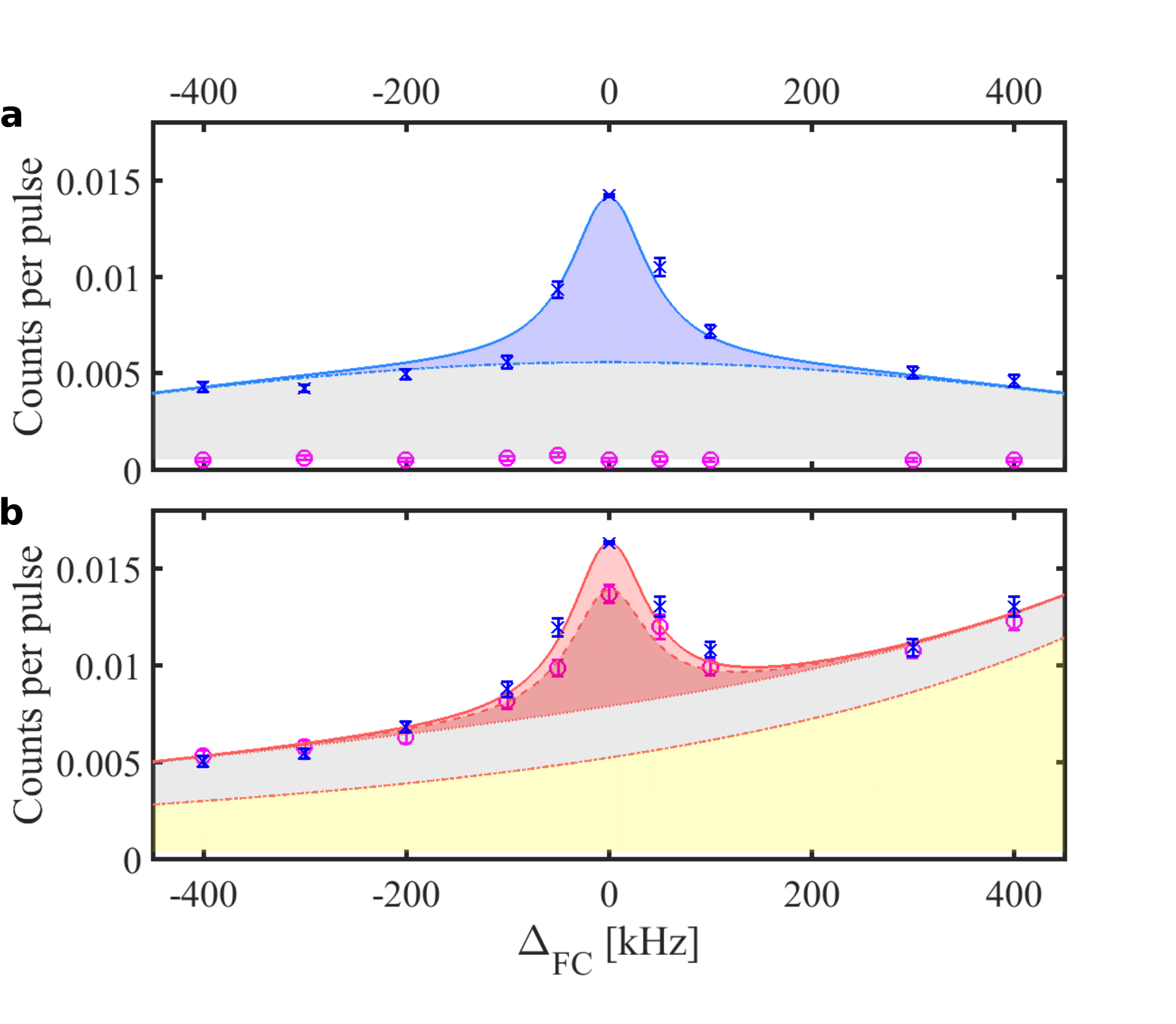}
	\caption{\textbf{Detected counts per pulse versus detuning of the filter resonance.} Zero detuning is one Zeeman splitting above (below) the write (read) excitation frequency. Each point represents around \num{1000} experiments with \num{55} repetitions each. The points on resonance with the write pulse include \num{60} times as many experiments.
		\textbf{a}, Heralding photon detection.
		\textbf{b}, Photon detection in the readout, considering only the first \SI{40}{\micro\second}. Blue crosses show data with write pulse present, magenta circles with write pulse off. The solid (dashed) lines show a fit with (without) write, containing scattered photons (blue, red), contribution from asymmetric excitations (grey), leakage (yellow) and background (unfilled).}
	\label{fig:FCavScan}
\end{figure}

\pagebreak 
\paragraph{Spectrum of scattered photons.}
First, we analyse the spectrum of the scattered photons by varying the resonance frequency of the spectral filter.
A weak write excitation pulse with a duration of about \SI{33}{\micro\second} is sent, and the photons transmitted through the filtering stages are detected (see fig.\ \ref{fig:FCavScan}a). The frequency of the scattered photons is blue-detuned by $\nu_Z$ with respect to the write excitation. We observe a narrow-band component associated with the symmetric Dicke state, above a broad background which is due to scattering associated with short-lived asymmetric excitations of the atoms. The width of the narrow peak is determined by the width of the spectral filter.
We define the write efficiency to be the ratio of these contributions, which is equal to $(63\pm1)\%$. It corresponds to the probability of having created a symmetric Dicke state upon detection of a scattered photon during the write process. The mean number of counts per pulse at zero detuning of \num{0.014} leads to \num{0.23} scattered photons per pulse in the cell cavity mode after correction for the detection efficiency and the escape efficiency out of the cell cavity. Counts from leakage of the excitation pulse are completely suppressed by polarization and spectral filtering. Further background counts are negligible during the write pulse.

A read pulse is sent after the end of the write pulse, with $\tau_\mathrm{D} = \SI{30}{\micro\second}$ and a similar energy. The frequency of the read pulse is blue-detuned by $2 \times \nu_Z$ from the write pulse such that the desired read-out photons have the same frequency as the heralding photons.
Scanning  the  filter resonance we observe a narrow peak above a broad background and an extra noise component (fig.\ \ref{fig:FCavScan}b). The narrow peak contains the retrieved photons, while the extra noise is due to the read excitation light leaking through the spectral filter.
Due to linear birefringence caused by detuning-dependent atom-light interaction, and different phase-shifts for the write and read excitation pulses arising from the temporal decay of the initial atomic state, the polarization filtering cannot be optimized for write and read excitation light simultaneously.

When repeating the same experiment without a write excitation pulse, we still observe a significant contribution in the number of read detection events. This is because the splitting of the two ground states is small compared to the detuning from the excited states, such that the read excitation field couples $\ket{g}$ and $\ket{s}$ via both the $\ket{m^\prime_F =3}$ (dashed transition in fig.\ \ref{fig:ExpSetup}d) and $\ket{m^\prime_F=4}$ excited manifolds with comparable strength. The read excitation thus creates atomic excitations through transitions from $\ket{m_F=4}$ to $\ket{m_F=3}$ and simultaneously reads them out by driving them back. This FWM process leads to short-term non-classical correlations, which, however, cannot be resolved with our setup and are mixed with the long-lived correlations generated by the write pulse. Hence, those otherwise interesting correlations \citep{Podhora2017,Zhu2017} have to be considered as noise here. 
When we include the write excitation pulse, we observe an increase in detection events from the readout of the excitation generated in the write step. We observe that this desired readout and the FWM noise are spectrally indistinguishable. Under the present conditions, the desired readout is weaker than FWM and leakage noise, while further backgrounds are negligible.

\pagebreak
\paragraph{Long-lived non-classical correlations.}
\label{CorrelationIntro}
In order to verify the quantum nature of the scheme we test for a violation of the Cauchy-Schwarz inequality $R=(g^{(2)}_{\mathrm{wr}})^2/(g^{(2)}_{\mathrm{ww}}g^{(2)}_{\mathrm{rr}})<1$ where subscripts $\mathrm{ww}$, $\mathrm{rr}$ refer to normalized second order auto-correlation functions for write and read fields and subscript $\mathrm{wr}$ to cross-correlation between the write field and the following read field \citep{Kuzmich2003}.
A nice feature of our system is that the single-photon wave packets have a long duration set by the inverse bandwidth of the filter cavity, which is much longer than the detector dead time.  This makes it possible to distinguish photon number states with a single detector. The correlation functions are then calculated from average number of counts according to $g^{(2)}_{{ij}} = \left< n_{i}(n_{j}-\delta_{ij})\right>/(\left<n_{i}\right>\left<n_{j}\right>)$ with $i,j$ $\in\left\{\mathrm{w,r}\right\}$ and ${n_\mathrm{w}}$ $({n_\mathrm{r}})$ is the number of detector clicks during the write (read) process. $\delta_{ij}$ is the Kronecker delta accounting for the non-commuting annihilation operators appearing in the auto-correlation functions.

\begin{figure}[ht!]
	\centering
	\includegraphics[width = 0.8\textwidth]{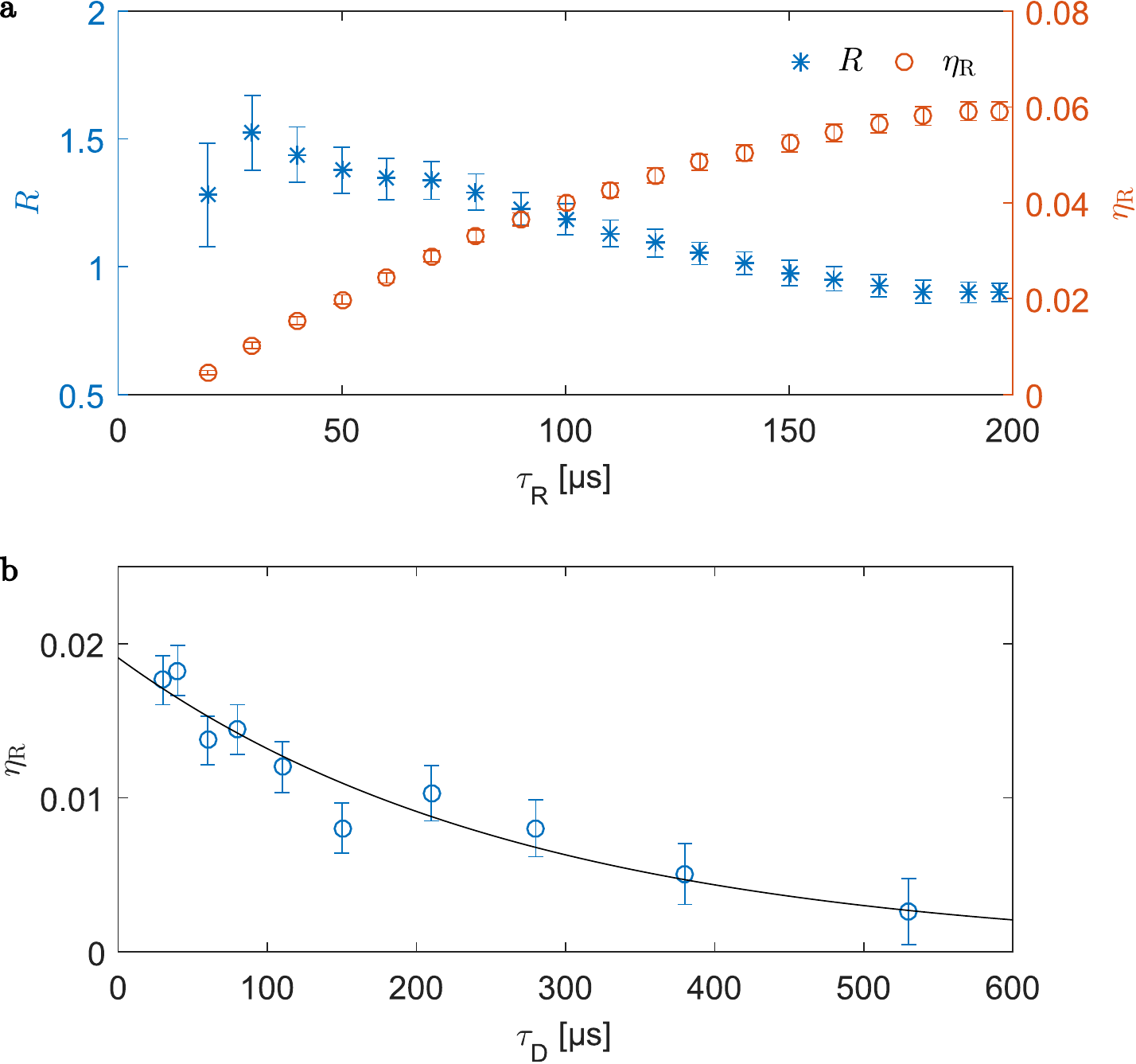} 
	\caption{\textbf{Temporal dynamics of the readout.}
		\textbf{a}, Cauchy-Schwarz parameter (blue asterisks, left axis) and retrieval efficiency (red circles, right axis) versus read detection integration time for $\tau_\mathrm{D}=\SI{30}{\micro\second}$. We observe violation of the Cauchy-Schwarz inequality for $\tau_\mathrm{R}< \SI{140}{\micro\second}$ while the retrieval efficiency increases throughout the read pulse. To limit the influence of noise we choose $\tau_\mathrm{R} = \SI{40}{\micro\second}$ for our correlation analysis.
		\textbf{b}, Retrieval efficiency versus write-read delay, for $\tau_\mathrm{R} = \SI{40}{\micro\second}$. An exponential fit (line) yields a $1/e$ collective-excitation lifetime of $\tau = 0.27\pm\SI{0.04}{\milli\second}$. We observe non-classical correlations for $\tau_\mathrm{D}<\SI{80}{\micro\second}$ (see Supplementary Information).
		The plotted values in the two graphs originate from different datasets hence the points at $\tau_\mathrm{D}=\SI{30}{\micro\second}$ and $\tau_\mathrm{R}=\SI{40}{\micro\second}$ are not identical. The top graph dataset corresponds to $\Delta_\mathrm{FC} = 0$ in fig.\ \ref{fig:FCavScan}.}
	\label{fig:ncnu&RvsTauRead}
\end{figure}

In the experiment we send read pulses with a \SI{200}{\micro\second} duration, and vary the integration time $\tau_\mathrm{R}$ for the read detection window.
We define the retrieval efficiency as $\eta_\mathrm{R} = \left\langle n_{\mathrm{r} |\mathrm{w}} \right\rangle - \left\langle n_\mathrm{r}\right\rangle$, the heralded readout probability subtracted by the unconditional readout probability. Fig.\ \ref{fig:ncnu&RvsTauRead} (a) shows how the trade-off between $R$ and $\eta_\mathrm{R}$ varies with $\tau_\mathrm{R}$.

In the following, we set $\tau_\mathrm{R}$ to only \SI{40}{\micro\second} in order to increase the signal-to-noise ratio. At $\Delta_{FC} = 0$ and for $\tau_\mathrm{D}=\SI{30}{\micro\second}$,
we observe $R = 1.4\pm0.1$, confirming the non-classicality of the scheme within four standard deviations.
For the same parameters, we measure $\eta_\mathrm{R} = (1.55\pm 0.08)\%$, leading to an intrinsic retrieval efficiency $\eta_\mathrm{R}^\text{i} = (16.1\pm 0.9)\%$ at the cell cavity output when correcting for the transmission loss and detector quantum efficiency.
For a pure two-mode squeezed state, expected in this type of protocols in the absence of noise, theory predicts thermal auto-correlation functions ($g^{(2)}_{ii}=2$), hence $g^{(2)}_\mathrm{wr}>2$ is required to violate the Cauchy-Schwarz inequality \citep{Kuzmich2003}. We find significantly lower auto-correlation values, $g^{(2)}_\mathrm{ww}=1.86\pm0.07$ and $g^{(2)}_\mathrm{rr}=1.45\pm0.05$, allowing us to achieve non-classicality with $g^{(2)}_\mathrm{wr}=1.97\pm0.05$. We attribute the reduced auto-correlations to leakage of the read drive pulse and mixing of two independent thermal processes in the write step.

We observe an increased value for the cross-correlation function $\tilde{g}^{(2)}_\mathrm{wr}=2.08\pm0.07$ when using only the last \SI{20}{\micro\second} of the write pulse. We attribute this to the shorter effective delay between write and read-out photons. However the reduced photon statistics in this case does not allow us to extend this analysis to all of our data.

In fig.\ \ref{fig:ncnu&RvsTauRead} (b) we show the decay of the retrieval efficiency as the write-read delay $\tau_\mathrm{D}$ increases. From an exponential fit we extract a memory lifetime of $\tau = \SI{0.27(4)}{\milli\second}$, which by far exceeds previously reported memory times at the single-photon level for room-temperature atomic vapour memories \citep{Bashkansky2012,Dou2017}.
The collective excitation lifetime is expected to be a half of the transverse macroscopic spin amplitude decay time, separately measured to be $T_2 =$ \SI{0.8}{\milli\second} (see Methods) and to be governed by spin relaxation due to wall collisions.

It should be duly noted that our implementation does not represent a single-photon source due to the excessive photon counts from FWM during the read pulse. When conditioning on detected heralding we observe a readout auto-correlation $g^{(2)}_{\mathrm{rr}  |\mathrm{w}}=1.3\pm0.2$.

\pagebreak
\paragraph{Temporal shape of the read out photons.}
To determine the nature and weight of the undesirable components limiting the fidelity of the read-out photons, we fit their temporal shape using a model adapted from \cite{Dabrowski2014}. According to the model, the detected read-out photons have two contributions: a desired part from the readout of the atomic excitations created during the write process, and the unwanted result of the FWM process depicted in fig.\ref{fig:ExpSetup}(d), present even in the absence of the write step. The photons scattered from $\ket{F^\prime,m^\prime_F = 3}$ to $\ket{s}$ are not resonant with the filtering cavities and are thus not detected. The photons scattered on the $\ket{F^\prime,m^\prime_F = 4}$ to $\ket{g}$ transition, however, are indistinguishable from the desired read photons and lead to spurious detection events spoiling the fidelity of the readout.
The model includes a noise offset comprising a constant term accounting for background and dark counts, and a power- and time-dependent term accounting for contamination from the drive leaking through the polarization and spectral filtering stages (see Supplementary Information).

\begin{figure}[ht!]
	\centering
	\includegraphics[width=1\textwidth]{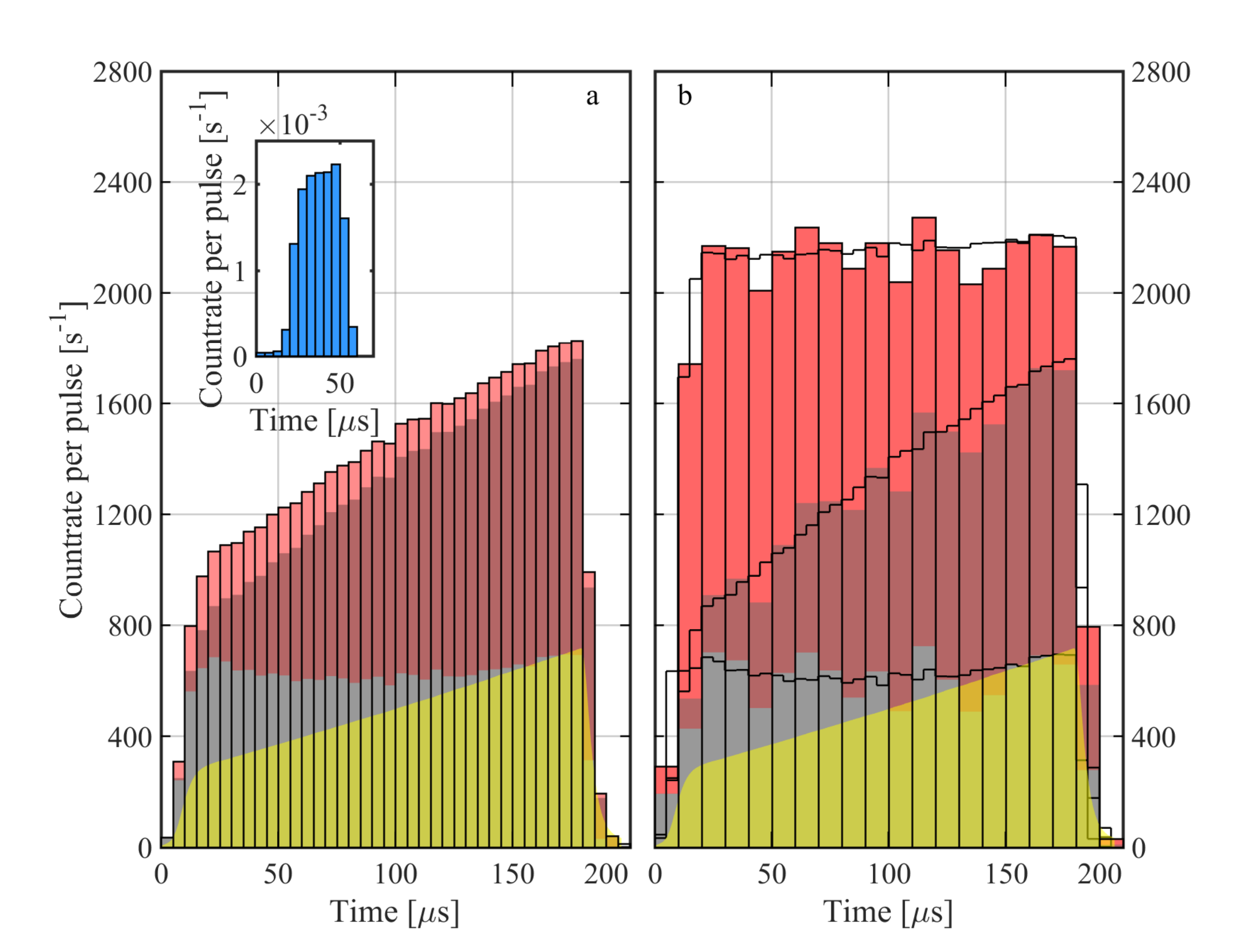}
	\caption{\textbf{Temporal shape of the detected read out photons.}
		\textbf{a}, Unconditional detection events.
		\textbf{b}, Heralded detection events. The bar graphs represent the detection events with a \SI{5}{\micro\second} (a), and \SI{10}{\micro\second} (b) binning. The inset represents the write photon with a \SI{5}{\micro\second} binning. In both plots the colored areas are obtained from a noise offset calibration and the fitted model. Yellow: noise offset. Dark red: FWM contribution. Red: atomic excitation readout. Grey: discrepancy between model and data. In (b), we replicate the data from (a) to guide the eye, using the number of collective excitations corresponding to the heralded case (black lines). The data presented corresponds to the point at zero detuning in fig.\ \ref{fig:FCavScan}. The origin of the horizontal axes are defined from the beginning of the respective detection windows.}
	\label{fig:PhotonShape}
\end{figure}

The temporal shape of the detected read out photons is shown in fig.\ \ref{fig:PhotonShape}, together with the fitted model.
The graphs show the normalized detected count-rates per time-bin during the readout process. Fig.\ \ref{fig:PhotonShape}\ (a) represents the unconditional detection events, while fig.\ \ref{fig:PhotonShape}\ (b) represents the heralded detection events conditioned on one or more write detection events, for the same dataset.
The values are normalized by the duration of the time-bins, and by the total number of pulses in (a) and by the number of trials with one or more write detection events in (b).
The total number of trials is \num{3248135}, and the total number of heralding events is \num{45774}. The temporal shape of the detected write photons is shown as an inset for reference.
In both graphs, the colored areas are obtained by using the same model with common coefficients for all the parameters except for the mean  number of collective excitations created during the write step, which we estimate from the number of write detection events and the total detection efficiency.
We attribute the discrepancy between the model and the data mainly to the fact that the model does not include FWM contributions from short-lived asymmetric collective excitations and does not account for decoherence of collective excitations.
See Supplementary Information for a description of the fit.

The signal-to-noise ratio rapidly starts to decrease with the integration time. The purity of the retrieved photons as well as the readout efficiency could be significantly higher if the FWM and leakage contributions can be suppressed.

\pagebreak
\section*{Outlook}
We have realized an efficient heralded photon source based on an atomic ensemble at room temperature, demonstrating a long single collective excitations lifetime of \SI{0.27(4)}{\milli\second} and a generation efficiency of $(63\pm 1)$\SI{}{\percent}. This lifetime could be extended significantly by employing a cell displaying a longer $T_2$ time.
We have demonstrated non-classicality of the light-matter correlations by observing a violation of the Cauchy-Schwarz inequality with four standard deviations. 
Even though the utility of those results is so far limited by excess noise from the leakage of the excitation light  and FWM, we emphasize that these do not present fundamental limitations.
The FWM contribution can be greatly reduced by using hyperfine instead of Zeeman storage, or suppressed by elaborate cavity design \citep{Nunn2017,Saunders2016}. Alternatively, the FWM can be eliminated in our setup by exciting the ensemble with circularly polarized light propagating along the magnetic field and storing the collective excitation in $\ket{F=4,m_F=2}$.  
The suppression of this two-photon transition for large detunings \citep{Vurgaftman2013} can be mitigated by using the caesium D1 line and an appropriate choice of detuning on the order of the excited state hyperfine splitting.
Further reduction of the remaining leakage will be possible by adding another filter cavity or by narrowing the filtering bandwidth. This narrowing will at the same time further improve the write efficiency. Finally, an active control of the polarization of the light at the cavity output could allow us to maximize the extinction of the polarization filter at all times.
With such improvements, our system could form the basis for a scalable room-temperature quantum repeaters.

\pagebreak
\section*{Methods}
\paragraph{Light.}
We use a home-built external cavity diode laser at \SI{852}{\nano\meter} that is locked to and narrowed in linewidth ($\leq \SI{10}{\kilo\hertz}$) by optical feedback from a triangular locking cavity. A slow feedback ($<\SI{}{\hertz}$) from a beatnote measurement of this excitation laser with a reference laser stabilized by atomic spectroscopy keeps the locking cavity resonance at a fixed detuning of \SI{925}{\mega\hertz} from the \num{4}-$\num{5}^\prime$ transition of the $D_2$ line of caesium. We send the excitation laser light through an acousto-optic modulator to pulse the light and choose the individual frequencies of the write and read pulses.

\paragraph{Vapour cell.}
The caesium vapour cell has a square cross section of $\SI{300}{\micro\metre} \times \SI{300}{\micro\metre}$ and a length of \SI{10}{\milli\metre}. It is coated with a spin-preserving anti-relaxation layer of paraffin (alkane). It is aligned along the optical axis of a low finesse ($\mathcal{F}\approx 18$) cavity  to enhance the light interaction. The losses of this "cell cavity" are dominated by the output coupler transmission. The vapour cell is inserted in a magnetic shield with internal coils that produce a homogeneous magnetic field perpendicular to the optical axis. We work at a Zeeman splitting frequency of $\nu_Z \approx \SI{2.4}{\mega\hertz}$
where the dissipated power in the coils heats up the vapour cell to around \SI{43}{\celsius}. Under these conditions we identify a coherence time of the ground state Zeeman levels of $T_2\approx \SI{0.8}{\milli\second}$, by performing a magneto-optical resonance spectroscopy measurement \citep{Julsgaard2004}.

\paragraph{Cavity stabilization.}
To stabilize the cell cavity length we input frequency-modulated light from the reference laser and derive an error signal from the transmitted signal. The error signal for the filter cavities are acquired from the transmission of frequency-modulated light from the excitation laser in the counter-propagating direction. The length of each cavity is then stabilized by a feedback acting on the respective piezo-actuated mirror mount. The lock light for the filter cavities is blocked by a chopper during the optical pumping and experiment periods.

\paragraph{Pumping.}
The atoms are initialized by circularly polarized pump and repump elliptical beams aligned along the magnetic field direction. The repump laser is locked on the $F=3$ to $F^\prime=2,3$ crossover of the $D_2$ line, the pump laser is locked on the $F=4$ to $F^\prime=4$ transition of the $D_1$ line. We typically observe an atomic orientation of $>  \SI{98.5}{\percent}$.

\paragraph{Filtering.}
We compensate for birefringence with a quarter wave plate and a half wave plate after the cell cavity and achieve a polarization filtering on the order of $10^{-4}$ with a Glan-Thompson polarizer. 
Spectral filtering is achieved by two concatenated triangular cavities. The first filter cavity is a narrow bandwidth cavity with a full width at half maximum (FWHM) of \SI{66}{\kilo\hertz} with an on resonance transmission of \SI{66}{\percent}. The second filter cavity has an on resonance transmission of \SI{90}{\percent} and a FWHM of \SI{900}{\kilo\hertz}. Both cavities together yield a spectral filtering of $7 \times 10^{-6}$ at a detuning of \SI{2.4}{\mega\hertz}. The cavities do not only provide filtering but also enable the motional averaging \citep{Borregaard2015}. They erase the 'which-atom' information by introducing a random delay due to the cavity photon lifetime. 

\paragraph{Detection efficiency.}
We measure the detection efficiency of the setup by sending a well-calibrated attenuated light pulse with the same polarization and frequency as the scattered photons through the system, and calculating the ratio of the count-rate versus the input rate obtained from the known power. We obtain a mean value of about \SI{9.6}{\percent} from the output of the cell cavity onto the single-photon detector (model COUNT-10C from LASER COMPONENTS), including the detector's quantum efficiency.

\paragraph{Cell cavity escape efficiency.}
We estimate the escape efficiency through the output coupler of the cell cavity from the transmission of this coupler ($\sim 20\%$) and the losses obtained from the finesse measurement. The obtained value is $\sim 62\%$.

\paragraph{Uncertainty estimation.}
To estimate the uncertainty of correlation functions we implement a bootstrapping technique. For a set of write and read pulses we obtain a distribution of the number of write and read counts in each write-read sequence. We then draw samples of the same size as the dataset from a probability distribution given by the dataset. For each sample we calculate the value of the correlation functions and as the number of samples increases, the variances of these bootstrap correlations converges. We find that the bootstrap correlations are close to normally distributed and the uncertainty estimates are given by the square root of the convergence values for the variances.

\section*{Acknowledgements}
The authors would like to thank M.\ Balabas for fabricating the cell used for this experiment.
The authors would also like to thank J.\ Borregaard for helpful discussions,
and acknowledge J.\ Appel and G.\ Vasilakis for their contributions at the early stages of the project.
Funding has been provided by the ERC AdG Interface, ERC consolidator QIOS, ARO grant W911NF, and by John Templeton Foundation.

\section*{Author contributions}
M.Z., K.B.D.\ and B.A.\ have contributed equally to the work, setting up and performing the experiment, analysing the data and writing the paper. A.S.S.\ and E.S.P.\ conceived the project. E.S.P.\ supervised the project.
All authors contributed in the writing of the manuscript.

\section*{Competing financial interests}
The authors declare no competing financial interests.

\pagebreak

\pagebreak
\renewcommand{\figurename}{Supplementary Figure}
\setcounter{figure}{0}
\section*{Supplementary Information}
\begin{figure}[ht!]
	\centering
	\includegraphics[width=0.8\textwidth]{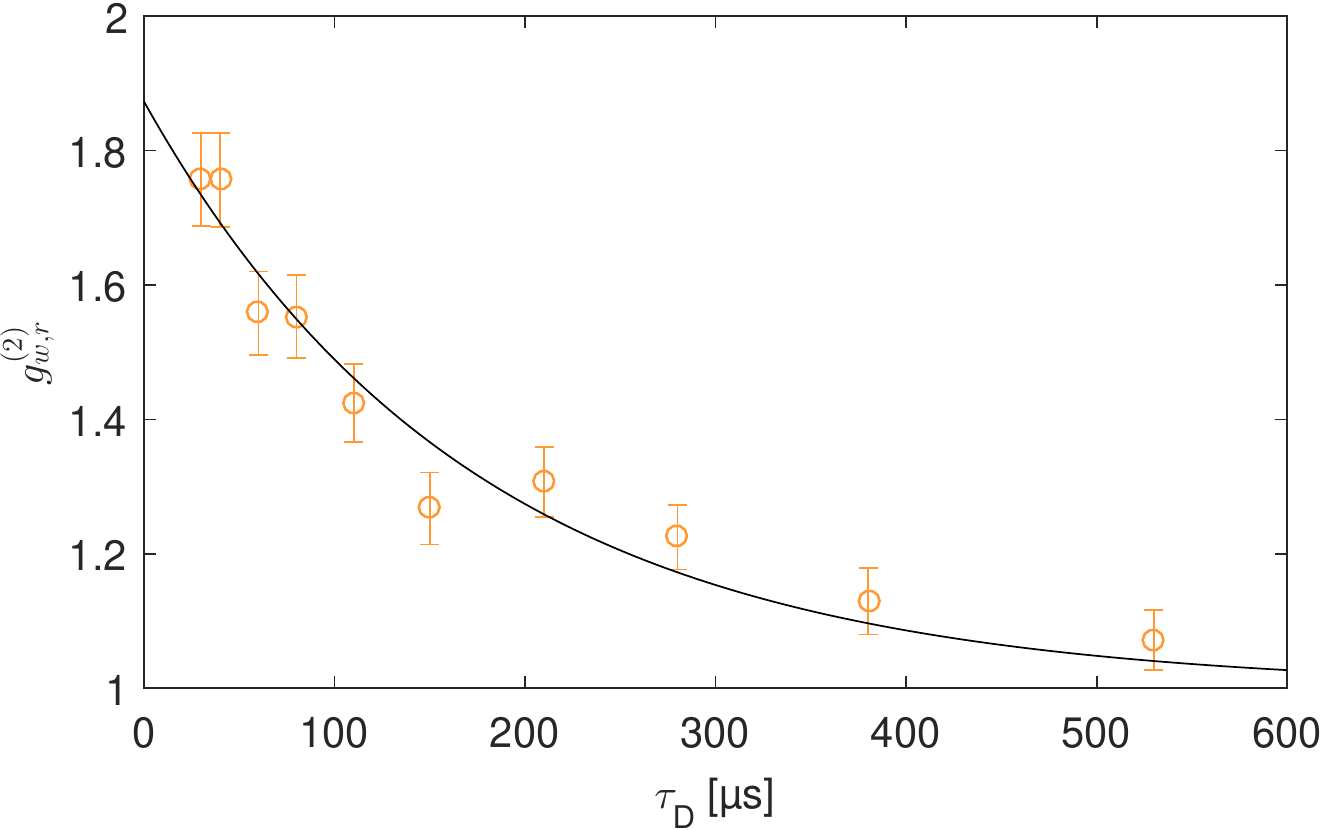} 
	\caption{Cross-correlation of write and read detection events versus write-read delay, for $\tau_\mathrm{R} = \SI{40}{\micro\second}$. A fit (line) to the function $g_{w,r}^{(2)}(t)=1+C\exp(-t/\tau_g)$  yields a characteristic decay time of $\tau_g=\SI{0.17(2)}{\milli\second}$.}
	\label{g2WR}
\end{figure}

\begin{figure}[ht!]
	\centering
	\includegraphics[width=0.8\textwidth]{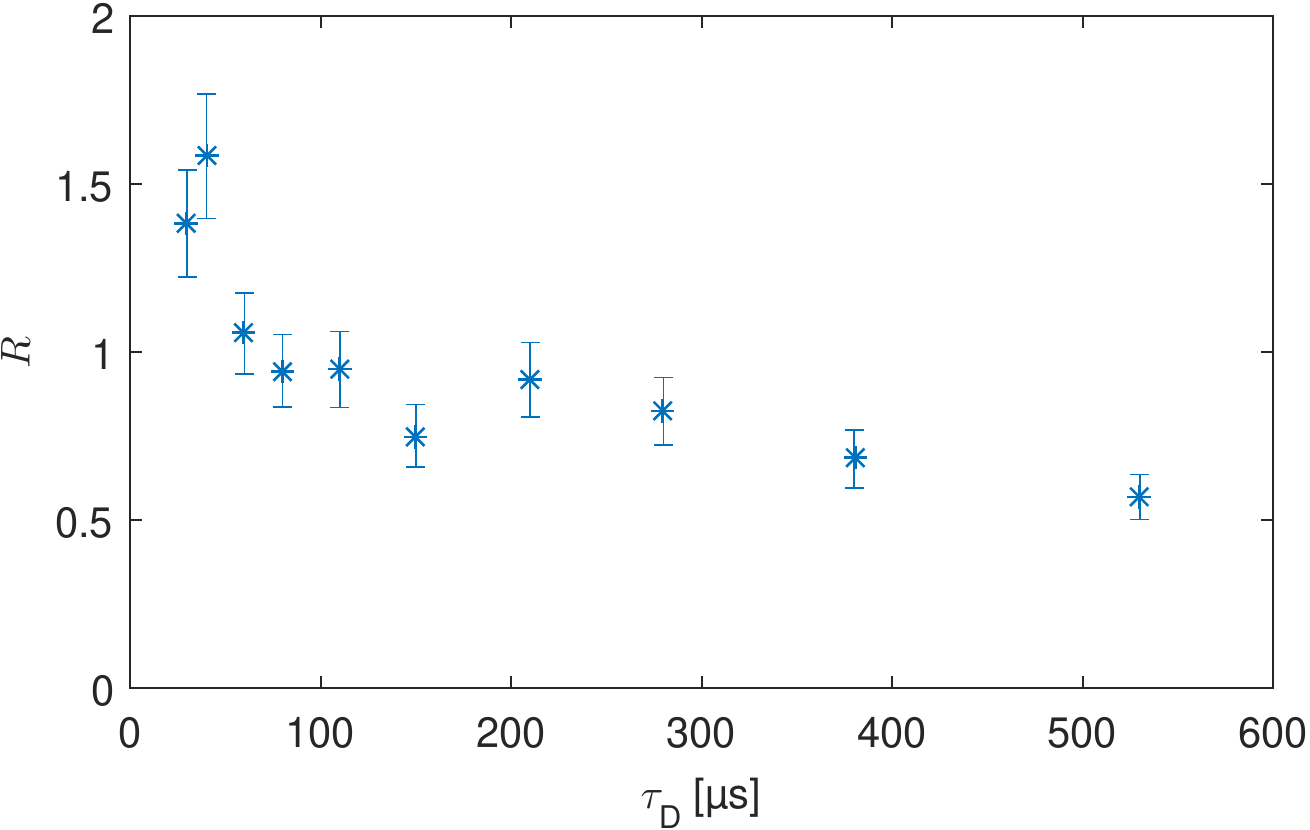} 
	\caption{Cauchy-Schwarz parameter decay versus write-read delay, for $\tau_\mathrm{R} = \SI{40}{\micro\second}$. We observe non-classical correlations ($R>1$) for $\tau_\mathrm{D}<\SI{80}{\micro\second}$.}
	\label{fig:CSR}
\end{figure}

\paragraph{Correlations decay.}
Supplementary figures \ref{g2WR} and \ref{fig:CSR} show the decay of the field correlations. We observe an increase in read leakage noise when the write-read delay increases. This increase is related to the cell cavity birefringence. This results in a faster decay for the observed cross-correlation than for the number of excitations. We can correct for the noise increase by considering the retrieval efficiency as it is done in the main text, thus getting a better estimate for the collective-excitation lifetime.


\paragraph{Read photon shape model.}
In order to model the temporal shape of the retrieved photons, we use the expression from \cite{Dabrowski2014supp}  for the mean number of detected photons as a function of time during the read process, adding a leakage and background contribution and accounting for atomic population decay:
\begin{align}\label{eq:PhotShapeModel}
\left\langle \hat{a}^\dagger_{ro}(t) \hat{a}_{ro}(t)\right\rangle = \chi_r^2(t)\Omega_R^2(t)\frac{N(t)}{N_0} \cdot \mathrm{exp} \left(t\left(\xi_r^2(t) - \chi_r^2(t)\right)\Omega_R^2(t)\frac{N(t)}{N_0}\right) \cdot n_{ce} & + \\\nonumber
\frac{\chi_r^2(t) \xi_r^2(t)\Omega_R^2(t)}{\xi_r^2(t) - \chi_r^2(t)}\frac{N(t)}{N_0} \cdot \left[\mathrm{exp} \left(t\left(\xi_r^2(t) - \chi_r^2(t)\right)\Omega_R^2(t)\frac{N(t)}{N_0}\right) - 1\right] & + \\\nonumber
L(t,\Omega_R^2(t)) + BG, &
\end{align}
where $\Omega_R$ is a common Rabi frequency defined for both transitions appearing in fig.\ 1 (d) in the article, $\chi_r$ and $\xi_r$ are reduced coupling coefficients, with the common Rabi frequency factored out.
$N_0$ is the number of atoms initially present in $\ket{g}$, and $N(t) = N_0 e^{-t/T_1}$ accounts for the atomic population decay with the time constant $T_1 \approx \SI{1.1}{\milli\second}$, which we measure independently.
$n_{ce}$ is the mean number of collective excitations created by the write pulse. 
$BG$ is a fixed offset given by background noise, mostly comprised of dark counts.
$L(t,\Omega_R^2(t)) = \left(L_0 + L_1 t \right) \Omega_R^2(t)$ is the leakage contribution, which is proportional to the intra-cavity power. Here the change in leakage due to birefringence dependent on the atomic population is approximated with a linear time dependency.
The mean number of photons per pulse after an integration time $\tau_R$ is then given by $\left\langle N_{ro}(\tau_R) \right\rangle = \int_0^{\tau_R} \left\langle \hat{a}^\dagger_{ro}(t) \hat{a}_{ro}(t)\right\rangle dt$.

The coefficients of the model are determined in the following way:
\begin{itemize}
	\item Background: 
	The background contribution $BG$ is determined from a part of the experimental sequence during which the SPCM gate is on and no light is sent to the system.
	\item Leakage: 
	In order to estimate the leakage contribution, we realize a fit identical to the one of fig.\ 2 in the article for consecutive time slices of \SI{32}{\micro\second} until the entire read detection window is covered. Each of the fits allows us to obtain a value for the leakage at zero detuning, and we then perform a linear interpolation of these values using $L(t,\Omega_R^2(t)) + BG$.
	\item Rabi frequency: 
	The common Rabi frequency $\Omega_R(t)$ is computed from the intra-cavity power of the cell-cavity, obtained via a calibrated photodiode monitoring the cell-cavity transmission and used to lock it.
	\item Number of collective excitations: 
	The number of collective excitations $n_{ce}$ is estimated from the mean number of write detection events corrected for the total detection efficiency including escape efficiency from the cell cavity, and by the write efficiency obtained from the data of fig.\ 2 in the article.
	\item Coupling coefficients:
	The ratio of the coupling coefficients $\alpha(t) = \frac{\xi_r(t)}{\chi_r(t)}$ is set by the Clebsch-Gordan coefficients at our detuning, 
	and by the cell cavity resonance affecting the coupling strength for the scattering of a photon on each side-band. The time dependency originates from the shift of the cell cavity resonance that is calibrated independently. The time-independent coupling coefficient $\chi_r$ is then determined from a fit as described below.
\end{itemize}

\begin{figure}[ht!]
	\centering
	\includegraphics[width=0.8\textwidth]{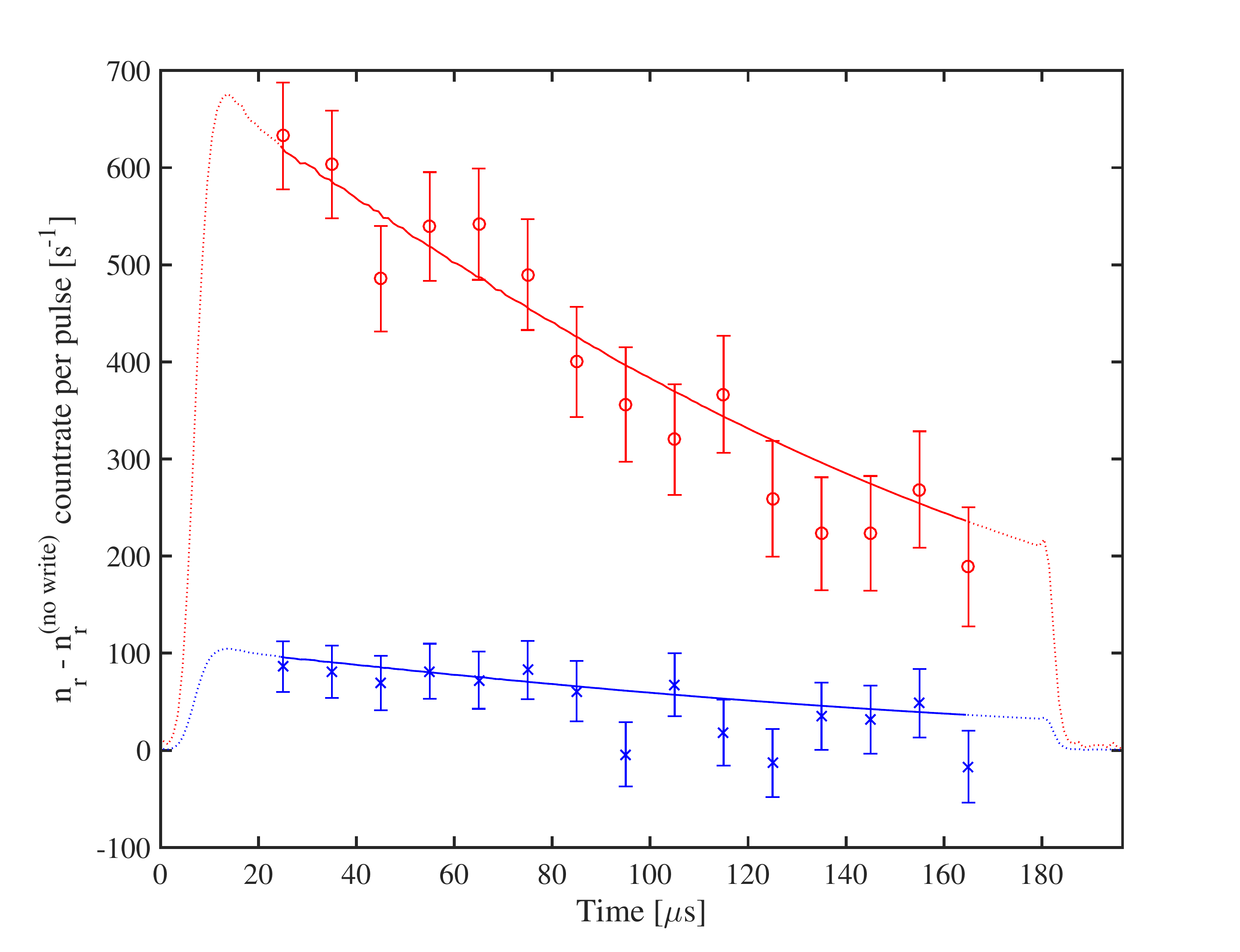} 
	\caption{\textbf{Read photon shape fit.}
		Blue: unconditioned data. Red: heralded data conditioned on at least one write detection event. In both cases the data without write has been subtracted. The markers represent the experimental data, and the lines are the fitted model. The solid part is the actual fitted region while the dotted part is an extrapolation using the amplitude of the drive pulses.}
	\label{fig:NReadDIff}
\end{figure}

We obtain the remaining parameter $\chi_r$ as the only free fit parameter from the difference of the read detection events with write and without write (suppl.\ fig.\ \ref{fig:NReadDIff}):
\begin{align}\label{eq:PhotShapeModelDiff}
n_r(t) - n_r^{no write}(t) = & \left\langle \hat{a}^\dagger_{ro}(t) \hat{a}_{ro}(t)\right\rangle - \left\langle \hat{a}^\dagger_{ro}(t) \hat{a}_{ro}(t) | n_{ce} = 0\right\rangle \\
= & \chi_r^2\Omega_R^2(t)\frac{N(t)}{N_0} \cdot \mathrm{exp} \left(t\left(\alpha^2(t) - 1\right)\chi_r^2\Omega_R^2(t)\frac{N(t)}{N_0}\right) \cdot n_{ce}. \nonumber
\end{align}

We then only have to consider the first term in eq.\ \eqref{eq:PhotShapeModel} and can disregard the leakage and background contributions.
We exclude the first and last \SI{25}{\micro\second} of the data for the fit in order to avoid influence of fast transients.
We fit both unconditioned data (suppl.\ fig.\ \ref{fig:NReadDIff}, blue) and heralded data (suppl.\ fig.\ \ref{fig:NReadDIff}, red) conditioned on at least one write detection event with common parameters. In the heralded case, we replace $n_{ce}$ by another value obtained using conditional probabilities.
We plug the obtained values into the original model, and plot the corresponding areas (explicit correspondency) in fig.\ 4 in the article, after low-pass filtering to account for the filter cavities buildup and ringdown effects.

We identify three possible reasons for the discrepancy between the data and the fit, corresponding to the grey area of fig.\ 4 in the article. Firstly, 
the cell cavity was not included in the Hamiltonian used to derive the model. Secondly, we neglect decoherence. However, the measured lifetime of \SI{0.27}{\milli\second} suggests that decoherence would likely have an effect on the timescale of the read detection window. Thirdly, no contribution from the fast-dephasing asymmetric collective excitations is considered. The latter we believe is the main cause for the discrepancy, but we lack sufficiently firm evidence to make stringent claims about this.

\end{document}